\documentclass{article}

\usepackage[utf8]{inputenc}
\usepackage{authblk}

\usepackage{graphicx}
\usepackage{placeins}
\usepackage{float}
\usepackage{subfigure}

\title{ResNets, NeuralODEs and CT-RNNs are Particular Neural Regulatory Networks}
\author{Radu Grosu}
\affil{Cyber Physical Systems\\ Technische Universit\"at Wien}
\date{}

\begin{document}
\maketitle

\begin{abstract}
This paper shows that ResNets, NeuralODEs, and CT-RNNs, are particular neural regulatory networks (NRNs), a biophysical model for the nonspiking neurons encountered in small species, such as the C.elegans nematode, and in the retina of large species. Compared to ResNets, NeuralODEs and CT-RNNs, NRNs have an additional multiplicative term in their synaptic computation, allowing them to adapt to each particular input. This additional flexibility makes NRNs $M$ times more succinct than NeuralODEs and CT-RNNs, where $M$ is proportional to the size of the training set. Moreover, as NeuralODEs and CT-RNNs are $N$ times more succinct than ResNets, where $N$ is the number of integration steps required to compute the output $F(x)$ for a given input $x$, NRNs are in total $M\,{\cdot}\,N$ more succinct than ResNets. For a given approximation task, this considerable succinctness allows to learn a very small and therefore understandable NRN, whose behavior can be explained in terms of well established architectural motifs, that NRNs share with gene regulatory networks, such as, activation, inhibition, sequentialization, mutual exclusion, and synchronization. To the best of our knowledge, this paper unifies for the first time the mainstream work on deep neural networks with the one in biology and neuroscience in a quantitative fashion.
\end{abstract}

\section{Introduction}
\label{sec:introduction}

Artificial neurons (ANs) were inspired by spiking neurons~\cite{bishop1995,gbc2016}. However, ANs are nonspiking and from a biophysical point of view, wrong. This precluded their adoption in neuroscience, from which they drifted apart. The main mistake of ANs was to combine in one unit the linear transformation of the inputs of a neuron, with a nonlinear transformation, inspired by  spiking. However, the nonlinear transformation itself is smooth, that is, it is nonspiking. 

Despite of this, the way in which an artificial neural network (ANN) puts the neurons together, corrects the mistake. In ANNs it is irrelevant what is the exact meaning of a neuron, and what is that of a synapse. What matters, is the mathematical expression of the network itself. This was best exemplified by ResNets, which were forced for technical reasons, to separate the linear transformation from the nonlinear one, and introduce new state variables, which are the outputs of the summation, and not of the nonlinear transformation~\cite{HeZRS15,heZR016}.

This separation, although not recognized as such in ResNets, allows us to reconcile ResNets with a biophysical model for nonspiking neurons, we call neural regulatory networks (NRNs), as they share many architectural motifs with gene regulatory networks. In particular, important binary motifs are activation, inhibition, sequentialization, mutual exclusion, and synchronization. As NRNs have a liquid time constant, we refer to them as LTCs in~\cite{icra2019}. NRNs capture neural behavior of small species, such as the C.elegans nematode~\cite{wicks1996dynamic}, and of nonspiking neurons in the the retina of large species~\cite{Kandel}. In this model the neuron is a capacitor, and its rate of change is the sum of a leaking current, and of synaptic currents. The conductance of synapses varies in a smooth nonlinear fashion with the potential of the presynaptic neuron, and this is multiplied with a difference of potential of the postsynaptic neuron, to produce the synaptic current. Hence, the smooth nonlinear transformation is the one that synapses perform, which is indeed the case in nature, and not the one of neurons.

The main difference between NRNs and ResNets, (augmented) NeuralODEs, and CT-RNNs~\cite{chen2018neural,dupont2019augmented,funahashi1993approximation}, is that NRNs multiply the conductance with a difference of potential. This is dictated by physics, as one needs to obtain a current. This constraint turns out to be very useful, as it allows NRNs to adapt to each particular input $x$. In other words, for each input, the discrete-time unfolding of an NRN is a ResNet, even though, there may be no ResNet that would properly work for all inputs. If the training set has size is $S$, this flexibility allows NRNs to be $M$ times more succinct than (augmented) NeuralODEs and CT-RNNs, where $M$ is proportional to $S$. Moreover, since (augmented) NeuralODEs and CT-RNNs are $N$ times more succinct than ResNets, where $N$ is the number of integration steps necessary to compute the output $y$ for a given input $x$, NRNs are in total $M\,{\cdot}\,N$ more succinct than ResNets. For a given approximation task, this considerable succinctness allows to learn a very small and understandable NRN, which can be explained in terms of the architectural motifs.

Another important difference of CT-RNNs and NRNs, compared to ResNets and (augmented) NeuralODEs, is the leaking current, which can be seen as a regularization term. If a neuron is not excited, the leaking current forces it to return to its equilibrium state, that is, it makes it stable. This regularization imposes beneficial restrictions when learning the parameters of an NRN.

The rest of the paper is structured as follows. In Section~\ref{sec:anns} we shortly review ANNs, and discuss their misconceptions. We then discuss ResNets in Section~\ref{sec:resnets} and show how they can provide a fresh view to ANNs. In Section~\ref{sec:neuralODEs} we review how one can smoothly transition from ResNets to NeuralODEs, and how approximation becomes a control problem. In Section~\ref{sec:ctrnns} we show that CT-RNNs are a slight extension of NeuralODEs. In Section~\ref{sec:NRNs} we show that all previous models are in fact a simplification of NRNs, and finally in Section~\ref{sec:conclusions} we draw our conclusions and discuss future work.

\section{Artificial Neural Networks}
\label{sec:anns}

\begin{figure*}[h]
    \centering
    \includegraphics[width=0.7\textwidth]{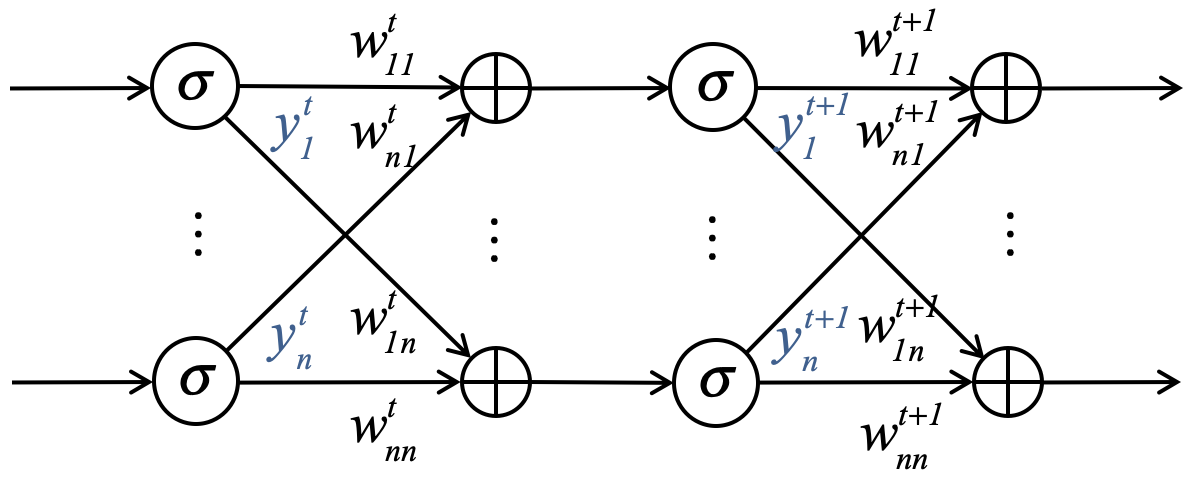}
    \vspace*{-3ex}
    \caption{The architecture of a DNN.}
    \label{fig:dnn}
    \vspace*{-1ex}
\end{figure*}

A main claim of ANNs is that ANs were inspired by their biological counterpart~\cite{bishop1995}. The story goes as follows: an AN receives one or more inputs (from other ANs), sums them up in a linear fashion, and passes the result through a non-linear thresholding function, known as an activation function. The latter usually has a sigmoidal shape. The threshold itself is a condition for the neuron to fire. 
Mathematically, this is expressed as follows:
\begin{equation}
y_i^{t+1} = \sigma(\sum_{j=1}^n w_{ji}^t\,y_j^{t}, \mu_i^{t+1})\qquad  \sigma(x,\mu) = \frac{1}{1+e^{x-\mu}} 
\end{equation}
where, as shown in Figure~\ref{fig:dnn}, $y_i^{t+1}$ is the output of neuron $i$ at layer $t\,{+}\,1$, $y_j^{t}$, for $j=1\,{:}\,n$, is the output of neuron $j$ at layer $t$, $w_{ji}^{t}$ is the weight associated to the synapse between neuron $j$ at layer $t$ and neuron $i$ at layer $t\,{+}\,1$, $\mu_i^{t+1}$ is the threshold of neuron $i$ at layer $t\,{+}\,1$, and $\sigma$ is the activation function, such as the logistic function above. A network with one input layer, one output layer, and $N\,{\geq}\,2$  hidden layers, is called a deep neural network (DNN)~\cite{gbc2016}.

An AN has unfortunately little resemblance to a biological spiking neuron, from which the thresholding idea originated, as its output is definitely not spiking. However, as we see in the later sections, ANNs are in fact very closely related to nonspiking neural networks. Let us look first at ResNets~\cite{HeZRS15}.

\section{Residual Neural Networks}
\label{sec:resnets}

DNNs with a very large number of hidden layers, were found to suffer from the so called degradation problem, which persisted even if the so called vanishing gradient problem was curated~\cite{HeZRS15,heZR016}. In short, the problem resulted from the inability of the DNNs to accurately learn the identity function. This manifested in the fact that a neural network with more layers, had poorer results, both in training and validation, than one with less layers. However, the first could have been obtained by simply extending the second with identity layers.

\begin{figure*}[h]
    \centering
    \includegraphics[width=0.7\textwidth]{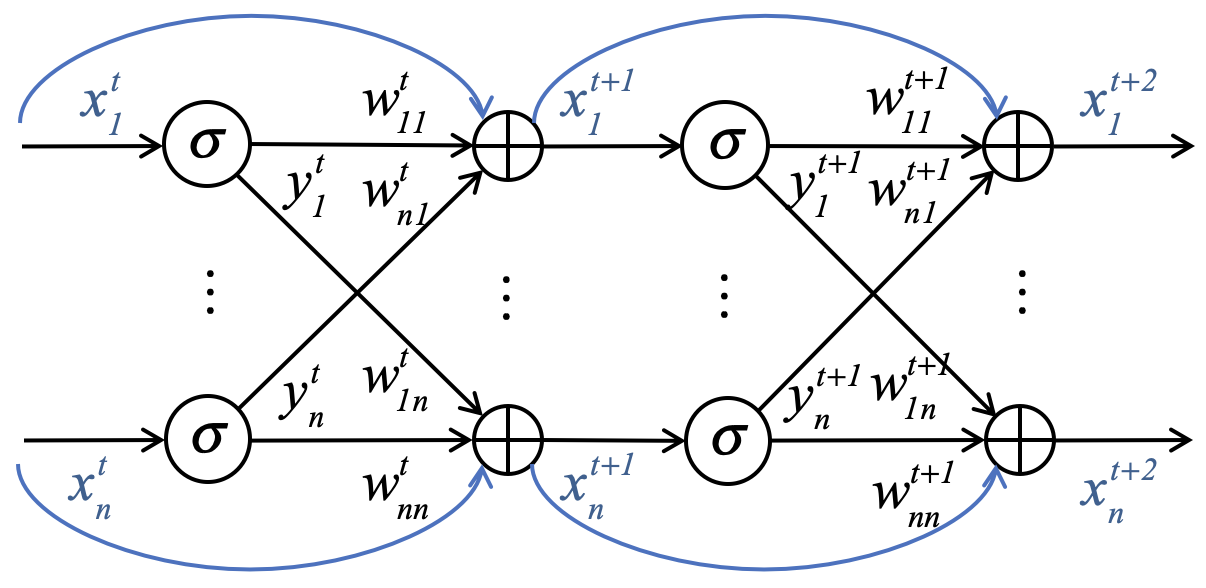}
    \vspace*{-3ex}
    \caption{The architecture of a ResNet.}
    \label{fig:resnet}
    \vspace*{-1ex}
\end{figure*}

Since identities were hard to learn, they were just added to the DNNs in form of skip connections. The resulting architecture, as shown in Figure~\ref{fig:resnet}, was called a residual neural network (ResNet)~\cite{HeZRS15,heZR016}\footnote{In~\cite{heZR016}, $x_i^t$ skips the first sum and it is added directly to $x_i^{t+2}$. Hence, the architecture shown in Figure~\ref{fig:resnet}, can be regarded as ResNets with finest skip granularity.}. In ResNets, the outputs $x_i^t$ of the sums are distinguished from the outputs $y_i^t$ of the sigmoids. Mathematically: 
\begin{equation}
x_i^{t+1} = x_i^t + \sum_{j=1}^n w_{ji}^t\,y_j^{t} \qquad y_j^{t} = \sigma(x_j^{t},\mu_j^t) 
\label{eq:resnet1}
\end{equation}
This distinction is very important from a biophysical point of view, although this was not explicitly recognized this way. The main idea is that neurons may just play the role of summation, and the sigmoidal transformation is a transformation happening in the synapses. In fact, one can put the weights in the synaptic transformation, too, which leads to the equivalent equations:
\begin{equation}
x_i^{t+1} = x_i^t + \sum_{j=1}^n y_{ji}^{t} \qquad y_{ji}^{t} = w_{ji}^t\,\sigma(x_j^{t},\mu_j^t) 
\label{eq:resnet2}
\end{equation}
Here $w_{ji}$ can be thought of as a maximum conductance of the input dependent synaptic transformation. These transformations are indeed in nature graded, that is nonspiking. As we will see later on, this observation is important to making the connection to biophysical models of nonspiking neurons. Before we further investigate this new idea, let us first look at NeuralODEs~\cite{chen2018neural}. 

\section{Neural Ordinary Differential Equations}
\label{sec:neuralODEs}

Equation~\ref{eq:resnet2}, can be simply understood as the Euler discretization of a differential equation, where the time step is just taken to be one~\cite{e17,chen2018neural}. Mathematically:
\begin{equation}
\dot{x}_i(t) = \sum_{j=1}^n y_{ji}(t) \qquad y_{ji}(t) = w_{ji}(t)\,\sigma(x_j(t),\mu_j(t)) 
\label{eq:contRS}
\end{equation}
In this equation, not only $x_i(t)$ and $y_i(t)$ change continuously in time, but so do the synaptic weights $w_{ji}(t)$. This might be not such a big problem, as during the discrete integration and optimization, one obtains ResNets anyway, and their weights can be learned. One just has to unfold the differential equation $N\,{=}\,T/dt$ times, where $T$ is the time horizon and $dt$ the integration step.

\begin{figure*}[h]
    \centering
    \includegraphics[width=0.85\textwidth]{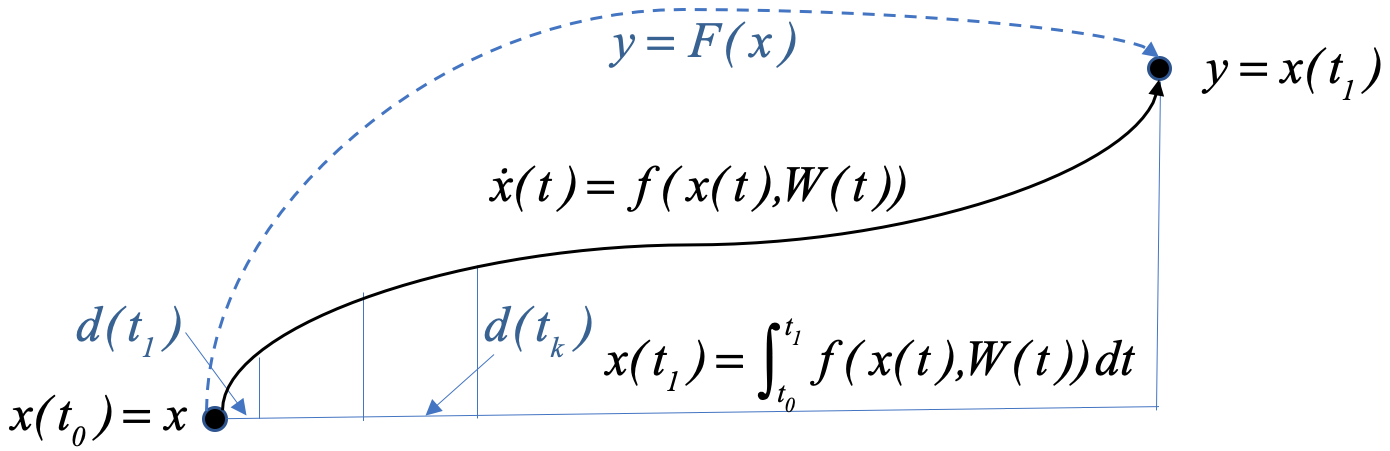}
    \vspace*{-3ex}
    \caption{Function approximation with ODEs.}
    \label{fig:odeApprox}
    \vspace*{-1ex}
\end{figure*}

At this point it is useful to make an excursion in function approximation with ODEs~\cite{e17}. Suppose we are given a set of data points $(x_i,y_i)$ for $i=1\,{:}\,S$, and that we would like to learn the function $y = F(x)$ which best approximates the data, and generalizes to data points not seen yet. ANNs with a single hidden layer directly learn $F$ with regression. They are universal approximators~\cite{cybenko1989approximation}.

However, if the approximation is learned with a ResNet in its differential form of Equation~\ref{eq:contRS}, then one actually learns the derivative $f$ of $F$, such that $\dot{x}(t)\,{=}\,f(x(t),W(t))$, $x(t_0)\,{=}\,x$ and $y\,{=}\,x(t_1)$. To approximate $F$ we thus need to compute the integral of $f$ from $t_0$ to $t_1$, as shown in Figure~\ref{fig:odeApprox}. In general, the derivative is simpler, and therefore learning DNNs is simpler, too.

ODE approximation can also be understood as a controllability problem~\cite{e17}: For any given position $x$, is it possible to find an input $W(t)$ for $f(x(t),W(t))$, that steers $x$ to $y$, within a finite horizon $T\,{=}\,t_1\,{-}\,t_0$?  The answer is affirmative, as ANNs with one layer are universal approximators, which implies that so are DNNs, and therefore ResNets, and consequently their ODE version, too.

Now suppose we make $W(t)$ time invariant, that is, $W(t)\,{=}\,W$. Are we still going to have a universal ODE approximator?  The answer is yes, as we will show in next section. In this case, the differential equations are as follows:
\begin{equation}
\dot{x}_i(t) = \sum_{j=1}^n\,y_{ji}(t) \qquad y_{ji}(t) = w_{ji}\,\sigma(x_j(t),\mu_j) 
\label{eq:neuralODEs}
\end{equation}
This is the form of Neural Ordinary Differential Equations (NeuralODEs)~\cite{e17,chen2018neural}. The $N$ times unfolding is still a ResNet but dramatically more succinct: it requires only $W$ parameters instead of $W\,{\cdot}\,N$! Could this work? This does indeed seem to be the case, with some slight extensions as discussed below. One intuitive explanation, is that modern numerical integrators would, for stability and efficiency reasons, use an adaptive time stepping d(t), as shown in Figure~\ref{fig:odeApprox}:
\begin{equation}
x_i(t+d(t)) = x_i(t) + \sum_{j=1}^n y_{ji}(t) \qquad y_{ji}(t) = w_{ji}\,d(t)\,\sigma(x_j(t),\mu_j) 
\label{eq:dtNeuralODE}
\end{equation}
Now the actual weights used are $W(t)\,{=}\,Wd(t)$, which are time dependent, and can vary depending on the stiffness of the network~\cite{e17}. 

NeuralODEs as approximators have the same dimensionality for the input, hidden state, and output. This is problematic, as ODEs do not allow trajectories to cross each other~\cite{dupont2019augmented}. However, adding sufficient dimensions, one can always avoid crossings. This can be achieved with the embedding of the input in the internal state, and a projection at the end of the state to the output, as follows:
\begin{equation}
x(t_0) = [x,0]^t\qquad y = \pi_S(x(t_1))
\label{eq:aNode}
\end{equation}
where the input is extended with zeroes as necessary, and the output is projected to the required subset. By applying such transformations, one arrives to what is called augmented NeuralODEs~\cite{dupont2019augmented}. Unfortunately, like NeuralODEs, augmented NeuralODEs are harder to train than ResNets, which is no wonder, as their size and parametrization is considerably more succinct. However, for learning, one can use the adjoint equation, and employ efficient numerical solvers~~\cite{e17,chen2018neural}.

\section{Continuous-Time Recurrent Neural Networks}
\label{sec:ctrnns}

One way to improve learning, is to add constraints to the approximation problem, or equivalently, to the control problem. Such constraints, can be understood as regularization constraints. Continuous-time recurrent neural networks (CT-RNN), introduce a stability regularization~\cite{funahashi1993approximation}. Their form is as follows:
\begin{equation}
\dot{x}_i(t) = -w_{i}x_i(t) + \sum_{j=1}^n\,y_{ji}(t) \qquad y_{ji}(t) = w_{ji}\,\sigma(x_j(t),\mu_j) 
\label{eq:ctrnn}
\end{equation}
The leading term $-w_{i}x_i(t)$ has the role to bring the system back to the equilibrium state, when no input is available. Hence, a small perturbation is forgotten, that is, the system is stable. This can be understood as a regularization constraint for the weights $W$ that have to be learned. Note that, except for this term, CT-RNNs are equivalent to NeuralODEs. 

CT-RNNs have been proven to be universal approximators, and the leading term does not play an important role in this proof~\cite{funahashi1993approximation}. Hence, NeuralODEs and augmented Neural ODEs are universal approximators, too.

\section{Neural Regulatory Networks}
\label{sec:NRNs}

Neural regulatory networks (NRNs), are a biophysical model for the neural system of small species, such as C.elegans~\cite{wicks1996dynamic,icra2019}, and of the retina of large species. Due to the small dimension of C.elegans, less than one millimeter, the neural transmission happens passively, in the analog domain, without considerable attenuation. Hence, the neurons do not need to spike for an accurate signal transmission. The distance between retina neurons is also very small.

In large species, spiking happens at the beginning of the axon of a neuron. Its role is to transform the amplitude of the analog signal, computed by the body of the neuron, in frequency, that is, in a train of spikes. The axon has a myeline stealth (an insulator for passive transmission), interrupted every millimeter by Nodes of Ranvier, which reinforce the signal. When the signal arrives at the synapse, it is converted back into a neurotransmitter concentration. Hence, the synapses work in an analog domain, too. Except for the digital transmission part, neurons and their synapses are therefore analog computation units. 

\begin{figure*}[h]
    \centering
    \vspace*{-1ex}
   \includegraphics[width=0.6\textwidth]{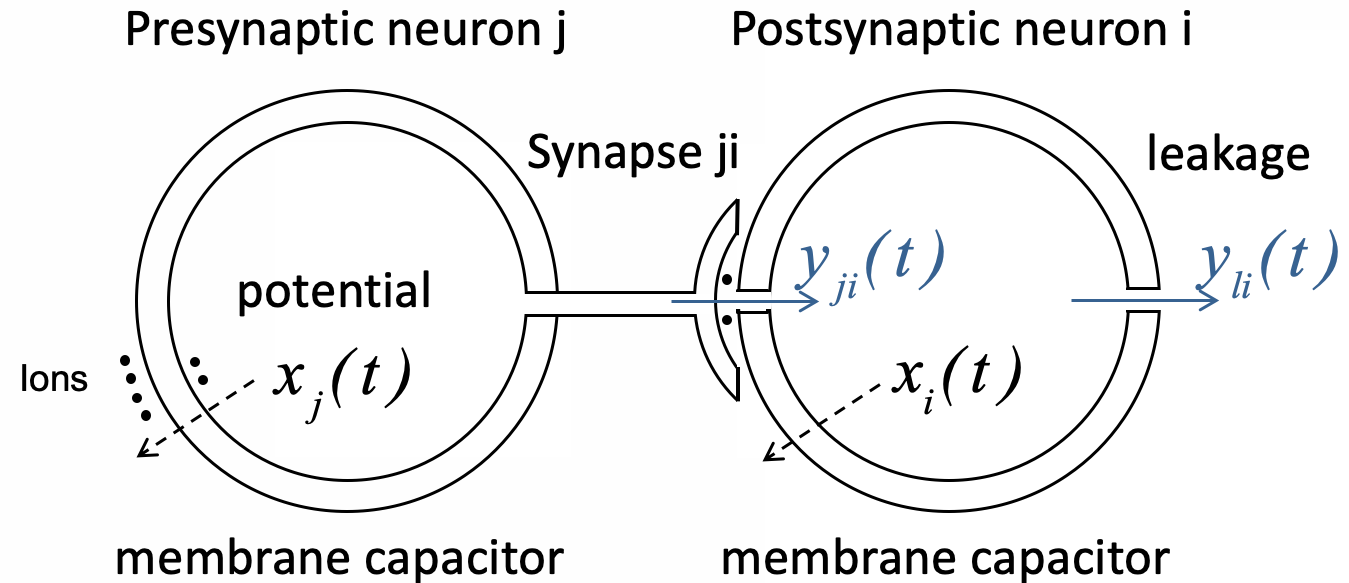}
    \vspace*{-2ex}
   \caption{The electric representation of a nonspiking neuron.}
    \label{fig:NRN}
    \vspace*{-1ex}
\end{figure*}

As shown in Figure~\ref{fig:NRN}, the membrane of a neuron is an insulator, with ions both on its outside and inside. Electrically, it plays the role of a capacitor. The difference of the inside and outside ionic concentrations defines the membrane potential $x$. The rate of change of $x$ depends on the currents $y$ passing through the membrane, which are either external currents (ignored here), a leakage current, and synaptic currents. For simplicity, we consider only chemical synapses. The capacitor equation, with capacitance $C\,{=}\,1$, is then as follows:
\begin{equation}
\dot{x}_i(t) = w_{li}\,(E_{li}-x_i(t)) + \sum_{j=1}^n y_{ji}(t) \quad y_{ji}(t) = w_{ji}\,\sigma(x_j(t),\mu_j)\,(E_{ji} - x_i(t)) 
\label{eq:NRN}
\end{equation}
where $E_{li}\,{=}\,-70$\,mV is the resting potential of the neuron, $w_{li}$ its leaking conductance, and $E_{ji}$ the synaptic potentials, 0\,mV in case of excitatory synapses (potential $x_j$ is negative so the current is positive), or -90\,mV for inhibitory synapsess (this is smaller than any potential so the current is negative).  

This equation is very similar to the CT-RNN equation. It has a leaking current, which ensures the stability of the network, and a presynaptic-neuron controlled conductance $\sigma$ for its synapses, with maximum conductance $w_{ji}$. This conductance is multiplied with a difference of potential $E_{ji}\,{-}\,x_i(t)$, to get a current. This biophysical constraint, makes it different from CT-RNNs.

So what is the significance of this last term for learning? One way to look at it, is that for some fixed paramterization $W$, the actual weight of the synapse from neuron $j$ to neuron $i$ is $w_{ji}(t)\,{=}\,w_{ji}\,d(t)\,x_i(t)$. Hence, it not only varies in time with the integration step $d(t)$, but also with the actual input $x$ provided to the network, and propagated as $x_i(t)$. In other words, an NRN generates for every particular input, a particular ResNet through unfolding, even though, there might be no single ResNet that would be appropriate for all inputs. 

This flexibility makes NRNs $M$ times more succinct than NeuralODEs and CT-RNNs, where $M$ is proportional to the size of the training set. As the latter are $N$ times more succinct than ResNets, NRNs are in total $M\,{\cdot}\,N$ times more succinct than ResNets. Hence, for a given approximation task, the learned NRNs are very small and explainable in terms of the architectural motifs, such as the binary motifs activation, inhibition, sequentialization, mutual exclusion, and synchronization, or more powerful higher order motifs. 

Another way of looking at $w_{ji}(t)\,{=}\,w_{ji}\,d(t)\,x_i(t)$, is that the importance of a synaptic weight $w_{ji}(t)\,{=}\,w_{ji}\,d(t)$ between neuron $j$ and neuron $i$, is weighted by the actual value $x_i(t)$ of neuron $i$. If the value $x_i(t)$ is low, the synapse is regarded as less important, and if the value $x_i(t)$ is high, as more important. 

Finally, the importance of the multiplicative factor $x$ can be explained in terms of interpretability~\cite{melis18}. Starting from a linear regression $f(x)\,{=}\,\theta^{t}x$, which is interpretable as a weighted sum of features in $x$, the authors suggest the generalization $f(x)\,{=}\,\theta(x)^th(x)$, where parameters vector $\theta(x)$ is allowed to depend on $x$, and $h(x)$ is a vector of higher-order features. In case of DNNs, $h(x)$ can be taken as identity, as each successive layer introduces higher level features. Moreover, the input dependent parametrization $\theta(x)$ becomes $w^t\sigma(x,\mu)$. But $f(x)\,{=}\,\sum_j w_{ji}\,\sigma(x_j,\mu_{j})\,x_i$ are precisely the synaptic currents of $\dot{x}_i$ in NRNs. They ensure robustness, too: small changes of $x$ lead to small changes of $f(x)$. This is not the case for DNNs, ResNets, NeuralODEs, and CT-RNNs. 


\section{Conclusions}
\label{sec:conclusions}

We have shown that ResNets, (augmented) Neural ODEs, and CT-RNNs are all a particular simplification of NRNs, a biphysical model for nonspiking neurons and their synapses. Such neurons are found for example in small species, such as the C.elegans nematode, or in the retina of large species. 

We have also shown that NRNs are $M\,{\cdot}\,N$ times more succinct than ResNets, where $M$ is a factor proportional to the size of the training set, and $N$ is a factor proportional to the number of integration steps required for computing the approximation. For a given approximation task, the learned NRN is thus very small, and explainable in terms of its architectural motifs. This gives considerable hope towards the stated goal of explainable machine learning.

A formal proof for the bound $M$ is currently under work, and so are empirical examples for the power of NRNs, on important and nontrivial taks, such as, lane keeping and image recognition.

\section*{Acknowledgements}
The author would like to thank his PhD students Sophie Gr{\"u}nba\-cher, Ramin Hasani, Mathias Lechner, and Zahra Babaiee, and his colleagues Daniela Rus, Scott Smolka, Jacek Cyranka, Flavio Fenton, Tom Henzinger, and Manuel Zimmer, for their feedback and fruitful discussions. This work was supported by the EU projects Semi4.0, Productive4.0, iDev4.0, Adaptness, and PersoRad, the Austrian doctoral school LogicCS, the infrastructure project CPS/IoT Ecosystem, and the US NSF project CyberCardia.

\bibliographystyle{abbrv}
\bibliography{references}

\end{document}